\documentclass[acs,prb,citeautoscript,superscriptaddress,preprint,onecolumn]{revtex4-2}

\usepackage{graphicx}
\usepackage{amsfonts,amsmath,amssymb,amsthm}
\usepackage{epstopdf}
\usepackage{upgreek,xspace}
\usepackage{chngcntr}
\usepackage[colorlinks,allcolors=blue]{hyperref}
\usepackage[version=3]{mhchem} 

\newcommand{\Tc}{$T_\mathrm{c}$}

\newcommand{\Hc}{$H_{c2}^{||}$}
\newcommand{\fig}[1]{Fig.~\ref{Fig#1}}

\begin{document}

\bibliographystyle{apsrev}
\title{Pair-Breaking and Dimensionality in Spin-Orbit Coupled Superconductors}

\author{Reiley~Dorrian}
\affiliation{Department of Applied Physics and Materials Science, California Institute of Technology, Pasadena, California 91125, USA.}
\affiliation{Institute for Quantum Information and Matter, California Institute of Technology, Pasadena, California 91125, USA.}

\author{Mizuki~Ohno}
\affiliation{Department of Applied Physics and Materials Science, California Institute of Technology, Pasadena, California 91125, USA.}
\affiliation{Institute for Quantum Information and Matter, California Institute of Technology, Pasadena, California 91125, USA.}

\author{Elena~Williams}
\affiliation{Department of Applied Physics and Materials Science, California Institute of Technology, Pasadena, California 91125, USA.}
\affiliation{Institute for Quantum Information and Matter, California Institute of Technology, Pasadena, California 91125, USA.}

\author{Adrian~Llanos}
\affiliation{Department of Applied Physics and Materials Science, California Institute of Technology, Pasadena, California 91125, USA.}
\affiliation{Institute for Quantum Information and Matter, California Institute of Technology, Pasadena, California 91125, USA.}

\author{Joseph~Falson}
\email{falson@caltech.edu}
\affiliation{Department of Applied Physics and Materials Science, California Institute of Technology, Pasadena, California 91125, USA.}
\affiliation{Institute for Quantum Information and Matter, California Institute of Technology, Pasadena, California 91125, USA.}

\begin{abstract}



The response of ultra-thin superconducting materials under parallel magnetic fields is often leveraged to obtain insight into the nature of the condensate, including features attributable to unconventional forms of pairing. Despite there being multiple competing mechanisms responsible for suppressing superconductivity, it is common for these analyses to overlook certain depairing channels. Here we report an analysis of thickness dependent superconductivity in thin films of \ce{LaBi2} using the multi-mechanism Kharitonov-Feigel'man framework . By resolving field-enhanced superconductivity in the thin-limit, we obtain an estimate the role of spin exchange scattering, in addition to paramagnetic and orbital effects. Our analyses offer insight into how fundamental quantities such as the critical temperature as well as Pauli limit are defined, recasting the landscape for how scattering times in two-dimensional superconductors can be interpreted.


\end{abstract}

\maketitle

Within the conventional theory of superconductivity (SC) laid out by Bardeen, Cooper and Schrieffer (BCS) in the mid-20th century, the SC ground state is comprised of spin-singlet Cooper pairs which are susceptible to various pair-breaking mechanisms under an external magnetic field. In the field of two-dimensional superconductors, it is common to observe an enhancement of the zero-temperature upper critical field \Hc($T\to0$) applied parallel to the plane beyond the conventional Chandrasekhar-Clogston or Pauli limit $H_\text{P}^\text{BCS}=1.86\times$\Tc~stemming from weakly coupled BCS theory \cite{xi_ising_2016, devarakonda_clean_2020, lu_evidence_2015, rhodes_enhanced_2021, liu_type-ii_2020, de_la_barrera_tuning_2018, xie_ising_2025, fatemi_electrically_2018, lu_full_2018, Kim_STO_2012, poage_violation_2025, liu_interface-induced_2018, chen_two-dimensional_2025, zhang_spinorbit_2023,nam_Pb_16,falson_type-ii_2020}. This was first explained by Klemm, Luther and Beasley (KLB) as a consequence of strong spin-orbit scattering which effectively randomizes electrons' spin orientations and thereby weakens the paramagnetic effect of an applied field \cite{klemm_theory_1975} in the dirty-limit ($\xi/\ell\gg1$, where $\xi$ is the superconducting coherence length and $\ell$ the mean-free-path), yielding,

\begin{equation}
    \ln\left(\frac{T_\mathrm{c}}{T_{\mathrm{c},H=0}}\right)=\psi\left(\frac{1}{2}\right)-\psi\left(\frac{1}{2}+\frac{3\tau_\mathrm{SO}{H_\mathrm{c2}^{||}}^2}{4\pi T_\mathrm{c}}\right),
\end{equation}
\noindent
where $\psi$ is the digamma function, $T_{\mathrm{c},H=0}$ is the zero-field critical temperature, $T_\mathrm{c}$ is the critical temperature corresponding to upper critical field \Hc, and $\tau_\mathrm{SO}$ is the spin-orbit scattering time. Note that in this letter, factors of physical constants $\mu_B,\,\hbar,\,k_\mathrm{B},\,m_\mathrm{e},\,e$ and $c$ are implicit. It is important to recognize that this model fails to capture interlayer coupling effects in the form of an orbital depairing contribution and $\tau_\mathrm{SO}$ emerges as the only sample-dependent fitting parameter. Generalizing beyond this limited framework is possible by characterizing multiple pair breaking channels semi-classically by scattering rates, $\nu_i$. Remaining in the dirty limit under a parallel field $H$ and allowing for finite magnetic disorder, three such scattering rates are of interest; the orbital pair-breaking goes as
\begin{equation}
\nu_\mathrm{orb.} \propto \tau_\mathrm{tr}(p_\mathrm{F}t)^2{H}^2,
\end{equation}
\noindent
where  $\tau_\mathrm{tr}$ is the transport scattering time, $p_\mathrm{F}$ is the Fermi momentum, and $t$ is the sample thickness \cite{tinkham_1996}; the paramagnetic scattering rate is given as
\begin{equation}
\nu_\mathrm{para.} \propto \tau_\mathrm{SO}(H-n_\mathrm{S}J\langle S_\mathrm{z}\rangle)^2
\end{equation}
\noindent
where $\tau_\mathrm{SO}$ is the spin-orbit scattering time and the offset term containing the concentration of magnetic impurities $n_\mathrm{S}$, the exchange coupling $J$, and the expectation value of the magnetic impurity spin in the z direction $\langle S_\mathrm{z}\rangle$ represents the Jaccarino-Peter compensation effect \cite{jaccarino_ultra-high-field_1962}; finally, the magnetic scattering rate is given by
\begin{equation}
\nu_\mathrm{s} \propto N_\mathrm{F} n_\mathrm{S}J^2 S(S+1).
\end{equation}
which contains the normal state density of states $N_\mathrm{F}$, and the spin of the impurity $S$ \cite{abrikosov_contribution_1960}. All three mechanisms are included in the expression developed by Kharitonov and Feigel'man (KF) for \Tc ~\cite{kharitonov_enhancement_2005}, which has been applied to understand field-induced superconductivity in magnetically-doped thin films of amorphous Pb \cite{pb_fieldinduced_17} and more recently crystalline \ce{LaSb2} \cite{llanos_field_2026}.

Although it predates the KF framework by several decades, the KLB model can be thought of as a reduction of the KF model in the limit of negligible interlayer coupling ($\nu_\mathrm{orb.}\to0$) and no magnetic impurities ($\nu_\mathrm{s}\to0$). As the impact of magnetic scattering is not always obvious and potentially due to the mathematical complexity of the KF model, KLB remains a prominent tool for modeling the critical field behavior of ultra-thin SC with strong spin-orbit coupling (SOC). The spin-orbit scattering time $\tau_\mathrm{SO}$ extracted from KLB, compared with $\tau_\mathrm{tr}$ from normal-state magnetotransport, is then a standard figure-of-merit for determining the underlying origin of the enhanced \Hc, whether that be the conventional scattering picture or more exotic scenarios such as inhomogeneous SC similar to the Fulde-Ferrell-Larkin-Ovchinnikov state in ultra-thin Pb films \cite{Pb_2013} or Zeeman-like spin splitting in transition metal dichalcogenide Ising superconductors \cite{zhou_ising_2016, zhang_ising_2021}. An investigation into how this line of reasoning and the physical conclusions drawn from it are influenced by the simplifications of the KLB model is essential, yet has received little attention.

In this letter, we examine how scattering times obtained in an analysis of ultra-thin superconductivity with strong spin-orbit coupling vary when taking into account all pair-breaking channels and finite sample thickness. To this end, we employ thin films of \ce{LaBi2}, which was recently discovered to be synthesizable as high-quality crystals with a superconducting critical temperature of around \Tc$\sim$0.5~K \cite{dorrian_relativistic_2026}. The SOC enhanced growth mode enables our series to reach an ultra-thin limit of 2.1~nm while maintaining a superconducting ground state. Systematic analysis of superconducting transitions reveals a field-enhanced critical field behavior in the ultra-thin limit, plausibly attributable to the suppression of magnetic fluctuations in an applied field. These features allow us to quantify each scattering channel within the KF framework and compare the results with those obtained from KLB. We find a systematic discrepancy between the two models, with KLB yielding meaningful values only in the limit of zero thickness.
\begin{figure}[ht]
    \centering
    \includegraphics[width=85mm]{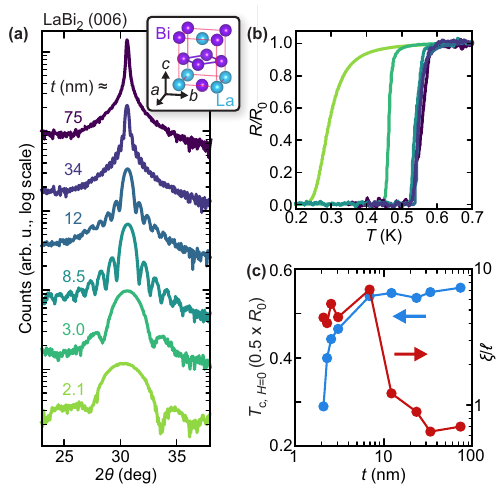}
    \caption{(a) X-ray diffraction of the (006) Bragg peak for several films across the thickness series in arbitrary units. Inset: schematic of the quintuple-layer building block which stacks to form the crystallographic structure of layered \ce{LaBi2}. (b) Zero-field superconducting transitions of the representative films, normalized by the normal-state resistance $R_0$ measured above the transition. (c) Superconducting critical temperature and the ratio of the superconducting coherence length ($\xi$) to the estimated mean free path ($\ell$) as a function of film thickness.}
    \label{Fig1}
\end{figure}

The \ce{LaBi2} thin-films were synthesized using molecular beam epitaxy on MgO (001) substrates according to the two-step growth procedure outlined in Ref~\cite{dorrian_relativistic_2026}. All films were capped \textit{in-situ} with amorphous Ge for protection from air exposure and measured within 24 hours of growth while being stored in an \ce{N2}-atmosphere glovebox. X-ray diffraction (XRD) (SmartLab, Rigaku) was performed to ensure phase purity of all films and estimate film thickness $t$ via analysis of Laue fringes (see Fig.~S1 of the Supplementary Material \cite{Supplement}). Electronic transport measurements of the superconducting state were gathered using a dilution refrigerator (base temperature $\sim$ 20 mK) equipped with a two-axis vector magnet (9--3 T). A linear four-point bond geometry was used on rectangular samples to measure the longitudinal resistance of the film, probed by a lock-in amplifier with a bias current of 100-500 nA at 10-20 Hz. To properly measure the in-plane critical field, a field alignment procedure was adopted from Ref.~\cite{llanos_field_2026}.

A full description of the KF model can be found in the Supplementary Material \cite{Supplement}. Omitting various proportionality constants and factors of \Tc, four independent variables are identified; $\nu_\mathrm{S}$, $\tau_\mathrm{SO}$, $\tau_\mathrm{tr}$, and the inverse Born parameter sgn$(J)\zeta=n_\mathrm{S}JS/\nu_\mathrm{S}$ (where $J$ is the exchange coupling strength and $S$ is the effective spin of the magnetic impurity). Notably, the $\tau_\mathrm{tr}$ term is weighted by a factor of $(p_\mathrm{F}t)^2$ where $p_\mathrm{F}$ is the Fermi momentum and $t$ is the thickness of the film; both factors are accounted for explicitly, acquiring an order-of-magnitude estimate of $p_\mathrm{F}$ using normal-state Drude analysis and the free-electron approximation (see Fig.~S2 of the Supplementary Material \cite{Supplement}). We also assume antiferromagnetic exchange coupling ($J>0$) and set $S=5/2$, following Ref.~\cite{llanos_field_2026}. $\nu_\mathrm{S}$ and the exchange coupling strength sgn$(J)\zeta$ are assumed to vary only slightly (allowed $\pm50\%$) across the thickness series. Finally, since $\tau_\mathrm{tr}$ should be the smallest scattering time of the system, we enforce a lower bound of $\tau_\mathrm{SO}/\tau_\mathrm{tr} \geq 1$ to restrict our fit parameters within physical limits.

The thickness progression of \ce{LaBi2} thin films from the ``bulk" to the ultra-thin limit is presented in \fig{1}. Single-phase XRD patterns are observed throughout the series, with strong Laue oscillations as seen about the dominant (006) Bragg peak in \fig{1}(a), indicating a sharp interface with the MgO substrate. The minimum thickness in the series is limited by the cohesion of \ce{LaBi2} islands which form in the first few minutes of deposition, without which the film lacks electrical continuity. Here we find this minimum thickness to be $t\approx 2.1$~nm. The structure of \ce{LaBi2} can be rendered in terms of stacked quintuple-layer blocks (QL, shown in the inset of \fig{1}(a)) which contain both corrugated \ce{La-Bi} bonds and \ce{Bi} square-nets \cite{llanos_monoclinic_2024,dorrian_stacking-selective_2025,kim_effect_2026,dorrian_relativistic_2026}. The thinnest films presented in this work are roughly 2.4 QL thick, or 1.2 conventional unit cells.

\begin{figure}[ht]
    \centering
    \includegraphics[width=85mm]{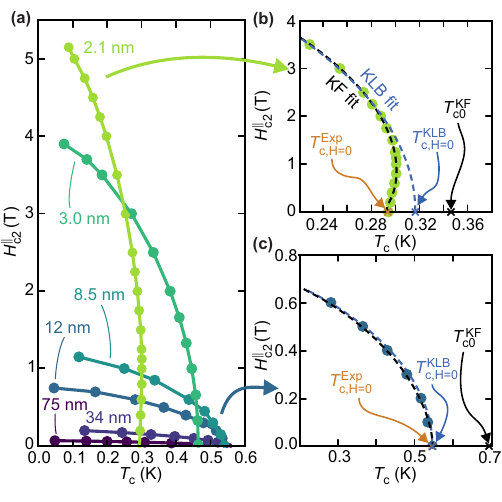}
    \caption{(a) In-plane upper critical field versus temperature for several film thicknesses. The lines shown here are an aid to the eye, not a quantitative fit. Zoom-in on the high-$T$ region for (b) $t = 2.1$ nm and (c) $t=12$ nm, showing fits to both the KF and KLB models. The three \Tc~definitions outlined in the text are marked on the horizontal axis.}
    \label{Fig2}
\end{figure}

Zero-field superconducting transitions across the series are presented in \fig{1}(b) and summarized in \fig{1}(c), where \Tc, defined as the temperature at which the resistance falls to half its normal-state value in zero applied field, is plotted as a function of film thickness $t$. \Tc~remains resilient over the first order-of-magnitude reduction in $t$ and then declines towards the 2 nm limit, a common observation in thin-film SC \cite{gubin_dependence_2005,semenov_optical_2009,chen_thickness_2000}. The estimated ratio of the superconducting coherence length ($\xi$) to the mean-free-path ($l$) (see Figs.~S2 and S4 of the Supplementary Material \cite{Supplement}) is presented in \fig{1}(c). We note that while our thinnest films satisfy the dirty-limit condition $\xi/\ell > 1$, the thicker films approach $\xi/\ell \sim 1$. Both the KLB and KF models are formulated in the dirty limit. Our application to the full thickness series therefore constitutes an approximation for the thicker films, where the models' quantitative accuracy may be reduced. The qualitative trends across the series, however, remain informative, and the key conclusions drawn from the ultra-thin limit---where the dirty condition is well satisfied---are robust.


The temperature-dependent critical field \Hc~for a field applied parallel to the film surface is shown in \fig{2} across the thickness series. In the thick limit ($(p_\mathrm{F}t)^2\gg1$), the orbital pair-breaking mechanism is dominant and results in relatively small parallel critical fields. This mechanism is suppressed with reducing $t$, leading to larger critical fields which increasingly surpass the Pauli limit. Using the zero-field \Tc~as defined in \fig{1} and extrapolating to $T=0$, the 2.1 nm film shows \Hc($T\to$0)/$H_\text{P}^\text{BCS}\approx 10$.

Close inspection of the \Hc($T$) curve for the 2.1 nm sample shows a departure from characteristics which can be captured by the single-mechanism KLB model, in the form of an enhancement of \Tc~for parallel fields as \Hc~is increased towards $\sim$1 T. There have been several reports of similar observations spanning a variety of microscopic interpretations \cite{asaba_WTe2_18,yang_magnetic_2026}. Intermediate transport scattering \cite{kogan_effects_2023} and interfacial Rashba SOC \cite{devizorova_interfacial_2024} are predicted to result in multiply-valued \Hc($T$) but have not yet been experimentally verified. A field-induced Fermi surface reconstruction has been discussed to explain re-entrant SC in gated \ce{LaTiO3-KTaO3} interfaces \cite{maryenko_re-entrant_2025}. Proximity to magnetic phases in heavy-fermion \cite{squire_superconductivity_2023} or iron-based SC \cite{hu_lithium_2025} often results in unconventional \Hc(T) behavior, and eavy Eu-doping in molybdenum chalcogenides \cite{meul_observation_1984} and infinite-layer nickelates \cite{vu_re-entrant_2026} leads to similar observations understood by the Jaccarino-Peter effect in which the ordering of magnetic dopants establishes a compensating exchange field. In our case, we note that a dilute concentration of magnetic contaminants (i.e. Ce, Nd, Fe) on the order of a few ppm is to be expected even in the high-purity La source material from AMES lab, used in this work. Building on previous work on isostructural Ce-doped \ce{LaSb2} \cite{llanos_field_2026}, we attribute the observed field-enhancement to the suppression of magnetic fluctuations by the polarizing force of an applied field, leading to an effective spin exchange scattering rate which is lower than the zero field value. This results in an anomalous positive slope in \Tc~as $H^\mathrm{\parallel}$ is ramped. We assume this mechanism to be dominant over the Jaccarino Peter effect in the low-density limit of magnetic impurities. Furthermore, magnetic scattering is neatly captured by the KF framework alongside other pair-breaking mechanisms, whereas alternative microscopic origins have not yet been incorporated into a tractable multi-mechanism model. For the sake of the comparative analysis pursued by this letter, we have elected to study the \Hc(T) curves using the KF framework.

Fits to both the KLB and KF are shown in \fig{2}(b) for $2.1$ nm and (c) $12$ nm, narrowed in on the low-$H$ region. Note that the KLB fit was performed for $H$ above the field-enhanced regime then extrapolated to zero-field. Full fit curves and corresponding parameter values across the thickness series are presented in the Supplementary Material \cite{Supplement}. We see the behavior in the ultra-thin limit is captured well by the KF model, while both models are qualitatively appropriate when the film thickness is made larger. The reason for this is the relative weight of depairing terms, where the orbital contribution begins to dominate in the thick films. However, here we wish to emphasize that this comparison invites ambiguity in the definition of the zero-field \Tc. We identify three distinct definitions; (i) the experimental $T_{\mathrm{c},H=0}^\mathrm{Exp.}$, that is the critical temperature measured via electronic transport at zero applied field; (ii) $T_{\mathrm{c},H=0}^\mathrm{KLB}$ as extrapolated to zero-field by fits to the KLB model; and (iii) $T_\mathrm{c0}^\mathrm{KF}$, the zero-field critical temperature in the absence of magnetic impurities as estimated by the KF model.

\begin{figure}[ht]
    \centering
    \includegraphics[width=85mm]{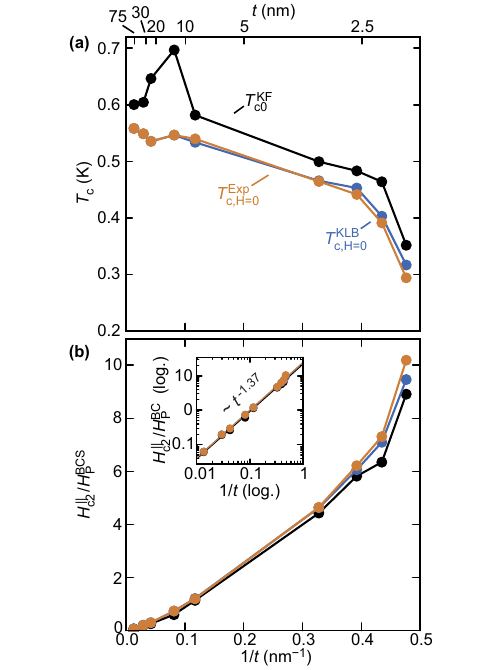}
    \caption{(a) Zero-field \Tc~ and (b) the ratio of the upper-critical field to the Pauli limit versus film thickness, for each of the three discussed interpretations of \Tc. (b, inset) $H_\mathrm{c2}^\parallel/H_\mathrm{P}^\mathrm{BCS}$ plotted on dual logarithmic axes, fitted to a linear model to extract the power-law scaling of $H_\mathrm{c2}^\parallel/H_\mathrm{P}^\mathrm{BCS}$ with film thickness.}
    \label{Fig3}
\end{figure}

\begin{figure}[ht]
    \centering
    \includegraphics[width=85mm]{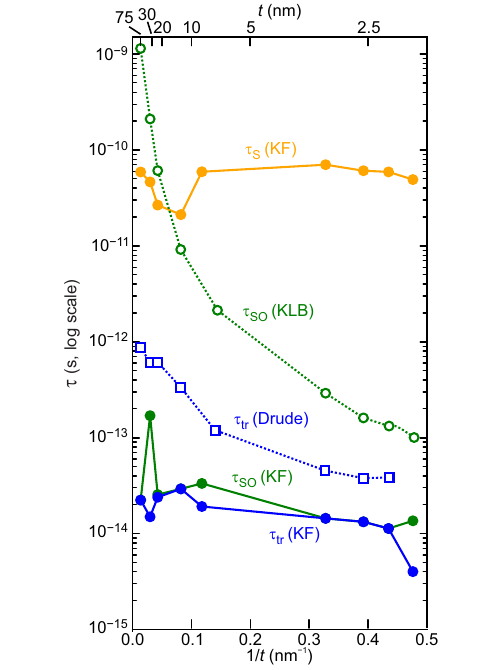}
    \caption{Characteristic scattering times relevant to pair-breaking as estimated by KLB, KF, and magnetotransport experiments, as a function of film thickness. }
    \label{Fig4}
\end{figure}

A summary of these three definitions of \Tc~is presented in \fig{3}a. For $t>5$~nm, $T_{\mathrm{c},H=0}^\mathrm{Exp.}$ and $T_{\mathrm{c},H=0}^\mathrm{KLB}$ agree well but begin to diverge as we approach the ultra-thin limit. Meanwhile, $T_\mathrm{c0}^\mathrm{KF}$ is found to be consistently larger than either $T_{\mathrm{c},H=0}^\mathrm{Exp.}$ or $T_\mathrm{c}^\mathrm{KLB}$ by about $\sim~10\%$, reflecting a background magnetic scattering rate which uniformly suppresses $T_\mathrm{c}^\mathrm{Exp.}$ even under zero-field. This analysis has implications on the definition of $H_\text{P}$, following the widely employed definition of $H_\text{P}^\text{BCS} = 1.86\times$\Tc. If \Tc~varies between analysis, so will $H_\text{P}$, as illustrated in Fig.\ref{Fig3}b using each approach to obtaining \Tc. The strong scaling of the observed violation of the Pauli limit with thickness $t$ is unsurprising given the strong impact of orbital scattering on \Hc which vanishes only in the strict two-dimensional limit. To fully suppress this orbital channel, we propose that the reported Pauli limit violation in any material should be an extrapolation to the thin limit, as parallel critical field comparisons at finite thickness inevitably conflate orbital and spin contributions. 
The inset of Fig.\ref{Fig3}b illustrates this by assuming a power-law scaling $H_\mathrm{c2}^\parallel/H_\mathrm{P}^\mathrm{BCS}\sim t^\alpha$ which we fit to our observed thickness-dependent violation ratio. We find $\alpha \approx -1.37$. 
Extrapolating this trend to the monolayer (single QL) limit, we would expect to find upper critical fields between roughly 26$\times$ and 30$\times$ the conventional Pauli limit in monolayer \ce{LaBi2}, depending on the chosen interpretation of \Tc.

We compare the thickness-dependent characteristic scattering times as estimated by various methods in \fig{4}. As mentioned above, the standard approach to obtaining $\tau_\mathrm{SO}$ is via the KLB model, while $\tau_\mathrm{tr}$ can be obtained by fitting normal-state magnetotransport data to the (single- or multi-carrier) Drude formula. This approach applied to the \ce{LaBi2} thickness series is summarized with open symbols in \fig{4}. The dramatic variation of $\tau_\mathrm{SO}$ (green) over four orders of magnitude is a systematic artifact from the KLB model's disregard for orbital pair-breaking. When the film thickness is non-negligible and the orbital mechanism suppresses the zero-temperature critical field, KLB is unable to distinguish this from an apparent reduction in spin-orbit scattering and hence significantly over-estimates $\tau_\mathrm{SO}$. Meanwhile, $\tau_\mathrm{tr}$ (blue) as estimated by Drude analysis is found to decrease by roughly an order of magnitude as the films approach the ultra-thin limit. This is reasonable in terms of the increase in structural disorder in ultra-thin films. We also note that, as discussed in the Supplementary Material \cite{Supplement}, $\tau_\mathrm{tr}$ here was estimated assuming an effective mass of $m_\mathrm{e}$ for both carrier types. Similar square-net semi-metals with conduction arising from overlapping $p$-orbitals host very light carriers with $m_\mathrm{eff}\sim0.05$-$0.1m_\mathrm{e}$ \cite{klemenz_topological_2019,hu_nearly_2017}. It is very likely, then, that the Drude-estimate $\tau_\mathrm{tr}$ presented in \fig{4} is over-estimated due to our treatment of effective mass.

We now compare the scattering times from KLB and Drude with those estimated by the KF framework, presented with solid symbols in \fig{4}. A finite magnetic scattering time ($\tau_\mathrm{S}=1/\nu_\mathrm{S}$, yellow) on the order of $10^{-11}$-$10^{-10}$ s is consistent with the behavior of all films in the series. This, along with the explicit treatment of film thickness, allows KF to more accurately disentangle the paramagnetic and orbital pair-breaking mechanisms. We find a significant distinction in the behavior of $\tau_\mathrm{SO}$ compared with the predictions of KLB, here largely independent of film thickness and falling generally within an order-of-magnitude of $\tau_\mathrm{tr}$. We attribute both observations to the acknowledgment of finite dimensionality by the KF framework, resolving the aforementioned artifact present in the results of the single-mechanism KLB theory.

Here it is worth cautioning the reader on the correlations between fit parameters in the KF model. While its ability to disentangle the orbital and paramagnetic mechanisms is greater than that of KLB, both pair-breaking channels share a similar quadratic field dependence which is only weakly offset by magnetic scattering effects, and the model becomes increasingly insensitive to the value of $\tau_\mathrm{SO}$ when $(p_\mathrm{F}t)^2\gg 1$. An estimate of $p_\mathrm{F}$ can be obtained from the Drude model, as discussed above, but in the case of materials like \ce{LaBi2} with complicated Fermi surfaces, even a multi-carrier fitting procedure can imperfectly reproduce magnetotransport data. Nevertheless, the $p_\mathrm{F}\propto n^{1/2}$ results in a linear dependence of $(p_\mathrm{F}t)^2$ on carrier concentration, but a squared dependence on $t$. Thus, we consider this fitting approach to be most suitable in cases where thickness series can be studied, including the thin-limit where the dimensionality term is rapidly suppressed. The scattering times obtained by this approach should be interpreted as qualitative indicators of the relative importance of each pair-breaking channel, rather than uniquely determined physical parameters. The impact of our previously-noted assumptions on the results of the fit is discussed in the Supplementary Material \cite{Supplement}.

In conclusion, we have targeted \ce{LaBi2} as a testbed for comparing models of depairing channels in thin two-dimensional superconductors, namely the KLB and KF frameworks. Neglecting the broader scope of pair-breaking mechanisms, as in KLB, can have a significant effect on the interpretation of the SC state. Important physical parameters such as the zero-field critical temperature and Pauli limit are not well defined when magnetic disorder is not taken into account. Lastly, the semi-classical scattering rates estimated by KLB are likely to be convoluted with the effects of finite sample thickness, which may be lifted by explicit consideration of orbital pair-breaking in the KF model. These results call for a reappraisal of how scattering rates from critical-field models are used to infer unconventional pairing mechanisms in two-dimensional superconductors.

We appreciate discussions with Mikhail Feigel'man, Yakov Fominov and Ding Zhang. This material is based upon work supported by the National Science Foundation Graduate Research Fellowship under Grant No. 2139433. We acknowledge funding provided by the Institute for Quantum Information and Matter, a NSF Physics Frontiers Center (NSF Grant PHY-2317110), and the support of JSPS Overseas Research Fellowship. 

\bibliography{bib}
\clearpage

\begin{center}
    \large{Pair-Breaking and Dimensionality in Spin-Orbit Coupled Superconductors\\
Supplementary Material}
\end{center}

\renewcommand{\thefigure}{S\arabic{figure}}

\setcounter{figure}{0}

\textit{The Kharitonov-Feigelman Model:}
We adapt the fitting procedure and KF model implementation from Ref.~[\onlinecite{llanos_field_2026}] with a few modifications for ease of computation. Under zero applied field, magnetic fluctuations from dilute spin-impurities suppress the measured $T_{\mathrm{c},H=0}$ below the intrinsic "un-doped" \Tc$_0$ via the Abrikosov-Gorkov formula \cite{abrikosov_contribution_1960}
\begin{equation}
    \ln\left( \frac{T_\mathrm{c0}}{T_{\mathrm{c},H=0}} \right) = \psi\left( \frac{1}{2} + \frac{\hbar\nu_\mathrm{S}}{2\pi k_\mathrm{B}T_{\mathrm{c},H=0}} \right)-\psi\left(\frac{1}{2}\right)
\end{equation}
where $\nu_\mathrm{S}$ is the magnetic scattering rate. Extending this to finite fields, the KF model predicts the suppression of \Tc~ below \Tc$_0$ via a sum over universal Matsubara frequencies $\varepsilon=2\pi T(n+1/2)$,
\begin{equation} \label{eqn:KF}
    \ln\left(\frac{T_\mathrm{c}}{T_{\mathrm{c}0}}\right)=\pi T\sum_\varepsilon\left(\frac{1}{|\varepsilon|}-C_0(\varepsilon)\right)
\end{equation}
where $C_0(\varepsilon)$ is the Cooperon Green's Function defined by the following expression,
\begin{equation}\label{eqn:KF2}
\begin{split}
    &\left( |\varepsilon| + \frac{1}{2}\nu_S \left[ 1+\frac{\langle S_\mathrm{z}^2\rangle}{S(S+1)} + \overset{\overset{\sim}{\wedge}}{L}_0(\varepsilon) - \delta\overset{\sim}{\Gamma}(\varepsilon) \right]\right. \\
     &\left. + \frac{3}{2}\tau_\mathrm{SO}\left[ h-\text{sgn(J)}\zeta\nu_\mathrm{S}\frac{\langle S_\mathrm{z}\rangle}{S}\right]^2+\frac{2}{9}(p_\mathrm{F}t)^2\tau_\mathrm{tr}h^2 \right) C_0(\varepsilon)=1
\end{split}
\end{equation}
Note that all factors of $\hbar$, $k_\mathrm{B}$, and $\mu_\mathrm{B}$ in Equation \ref{eqn:KF}-\ref{eqn:KF2} are implicit, and $h=\mu_\mathrm{B}H$ is implicitly $h/T_\mathrm{c0}$. The first bracketed term captures pair breaking via spin-dependent scattering off magnetic impurities which are polarized by an applied field. The operator $\overset{\overset{\sim}{\wedge}}{L}_0(\varepsilon)$ is defined as follows
\begin{equation}
    \overset{\overset{\sim}{\wedge}}{L}_0(\varepsilon)C_0(\varepsilon) = \nu_\mathrm{S}\frac{\langle S_\mathrm{z}^2\rangle}{S(S+1)}\cdot T\sum_\omega\frac{2\omega_\mathrm{S}}{\omega^2-\omega_\mathrm{S}^2}C_0(\varepsilon-\omega)
\end{equation}
where $\omega_\mathrm{S}$ is the Zeeman splitting, and $\delta\overset{\sim}{\Gamma}(\varepsilon)$ is defined in terms of the exchange scattering rate $\Gamma(\varepsilon)$ via $\delta\overset{\sim}{\Gamma}(\varepsilon) = 1-\Gamma(\varepsilon)/\nu_\mathrm{S}$. The second bracketed term captures the paramagnetic effect which is mediated by the spin-orbit scattering time $\tau_\mathrm{SO}$ and contains the effects of magnetic impurities establishing an exchange field which either enhances or counteracts the applied field $h$ (the Jaccarino-Peter effect). The final term, scaled by the transport scattering time $\tau_\mathrm{tr}$ and the dimensional factor $(p_\mathrm{F}t)^2$ where $p_\mathrm{F}$ is the Fermi momentum and $t$ the film thickness, represents the orbital pair-breaking mechanism. 

As noted in the main text, fitting to the KF model was first performed on the \Hc(T) data exhibiting field-enhanced $T_\mathrm{c}$ from the 2.1 nm sample. It is in this ultra-thin limit where (i) the dirty limit condition $\xi/\ell\gg 1$ is most satisfied, and (ii) the dimensional $(p_\mathrm{F}t)^2$ is minimized, allowing the model to better resolve the magnetic scattering and paramagnetic pair-breaking channels. Assumptions on the sign of the exchange coupling $J$ and the value of the impurity spin $S$ were made for consistency with Ref.~\cite{llanos_field_2026} (see Fig. S7 and Table II for the impact of these assumptions). The values of $\nu_\mathrm{S}$ and sgn$(J)\zeta$ obtained from this fit were used as the initial values for fitting the remainder of the series, with these parameters constrained to within $\pm50\%$ of this initial value. The initial guess for $\tau_\text{tr}$ was based on the estimate from Drude analysis, and $\tau_\text{SO}$ was initialized at $2\times\tau_\mathrm{tr}$ and restricted by the lower-bound $\tau_\mathrm{SO}/\tau_\mathrm{tr}\geq 1$. Full fit curves are presented in Fig. S6 and the values of all fit paramters used in the main text are presented in Table I.

\textit{The Klemm-Luther-Beasly Model:}
The Klemm-Luther-Beasly (KLB) model is formulated for a two-dimensional superconductor in the dirty limit ($\xi/\ell\gg 1$), giving
\begin{equation}\label{eq:KLB}
    \ln\left(\frac{T_\mathrm{c}}{T_{\mathrm{c},H=0}}\right)=\psi\left(\frac{1}{2}\right)-\psi\left(\frac{1}{2}+\frac{3\tau_\mathrm{SO}\mu_\mathrm{B}^2{H_\mathrm{c2}^{\parallel}}^2}{4\pi\hbar k_\mathrm{B}T_{\mathrm{c},H=0}}\right),
\end{equation}
where $T_{\mathrm{c},H=0}$ is the zero-field critical temperature, $\tau_\mathrm{SO}$ is the spin-orbit scattering time, and $H_\mathrm{c2}^\parallel$ is the applied parallel field. For a given $T_{\mathrm{c},H=0}$ and $\tau_\mathrm{SO}$, Eqn.~\ref{eq:KLB} was solved numerically to obtain $H_\mathrm{c2}^{\parallel}$ as a function of $T_\mathrm{c}$. This was then fit to the experimental \Hc(T) data with the free parameters $T_{\mathrm{c},H=0}$ and $\tau_\mathrm{SO}$. Full fit curves are presented in Fig. S6 and the corresponding fit parameters are listed in Table I.

\textit{Two-Carrier Drude Analysis:}
Fits of the magnetoresistance and Hall resistivity to the two-carrier Drude formulas, 
\begin{align}
    \rho_{xx} &= \frac{1}{e}\frac{(p\mu_h+n\mu_e)+(p\mu_e+n\mu_h)\mu_h\mu_eB^2}{(p\mu_h+n\mu_e)^2+(p-e)^2\mu_h^2\mu_e^2B^2}\\
    \rho_{yx} &= \frac{B}{e}\frac{(p\mu_h^2-n\mu_e^2)+(p-n)^2\mu_h^2\mu_e^2B^2}{(p\mu_h+n\mu_e)^2+(p-e)^2\mu_h^2\mu_e^2B^2}
\end{align}
were performed across the thickness series to extract estimated concentrations and mobilities for both hole- and electron-like carrier species. This is illustrated in Fig.~\ref{FigS2} for three representative thicknesses. In the thick-limit, non-quadratic deviations from expected Drude behavior in the MR leads to apparent discrepancies between the data and the two-carrier fits. In the ultra-thin limit, the emergence of cusp-like features attributed to weak anti-localization requires the fit to be performed within a truncated field region. For these reasons, all quantities extracted from Drude analysis in this work should be regarded only as approximate values. 

Rough estimates of mean free paths and transport scattering times were acquired from the expressions
\begin{gather}
    \ell_i = \frac{\mu_i}{e}p_{\mathrm{F},i} =\frac{\hbar\mu_i}{e}(3\pi^2n_i)^{1/3}\\   \tau_{\mathrm{tr},i} = \frac{m_{\mathrm{eff,}i}\mu_i}{e},
\end{gather}
where $i$ indexes the carrier type, $\ell_i$ and $\tau_{\mathrm{tr},i}$ denote the mean free path and transport relaxation time, respectively. Here, $\mu_i$ is the carrier mobility, $e$ is the elementary charge, $p_{\mathrm{F},i}$ is the Fermi momentum, $\hbar$ is the reduced Planck constant, $n_i$ is the carrier density, and $m_{\mathrm{eff,}i}$ is the effective carrier mass. Lacking an estimate of $m_{\mathrm{eff,}i}$ from, say, Shubnikov-de Haas oscillations or ARPES experiments, for simplicity we assume $m_{\mathrm{eff,}i} = m_\mathrm{e}$ for both carrier types,  where $m_\mathrm{e}$ is the free-electron mass. It is very likely that the true effective mass is much smaller as in other square-net semi-metals \cite{klemenz_topological_2019,hu_nearly_2017}, which would reduce $\tau_{\mathrm{tr},i}$ in proportion. Therefore, the discrepancy between the Drude and KF estimates of $\tau_{\mathrm{tr}}$ as shown in the main text may be resolved by a more thorough characterization of the carrier dynamics in \ce{LaBi2}.

For the purposes of the $(p_\mathrm{F}t)^2$ factor in the KF framework, the species-averaged carrier density $\langle n \rangle = \text{mean}(n,p)$ was fit to a power law relationship $\langle n \rangle\sim t^n$ excluding a couple of clear outliers. This extrapolation was used to estimate $p_\mathrm{F}$, such that the sample-to-sample variation was dominated by film thickness rather than potential errors in the Drude analysis.

It should also be noted that, due the experimental difficulty of repeatedly producing electrically conducting films at the minimum 2.1 nm thickness, we have only obtained normal-state magnetotransport data down to a slightly thicker 2.3 nm limit. It is assumed that this does not significantly effect the results of the Drude analysis, however for transparency we have omitted a Drude estimate of the transport scattering time for the 2.1 nm film in the main text.
\newpage

\newpage
\begin{figure}[h]
    \centering
    \includegraphics[width=170mm]{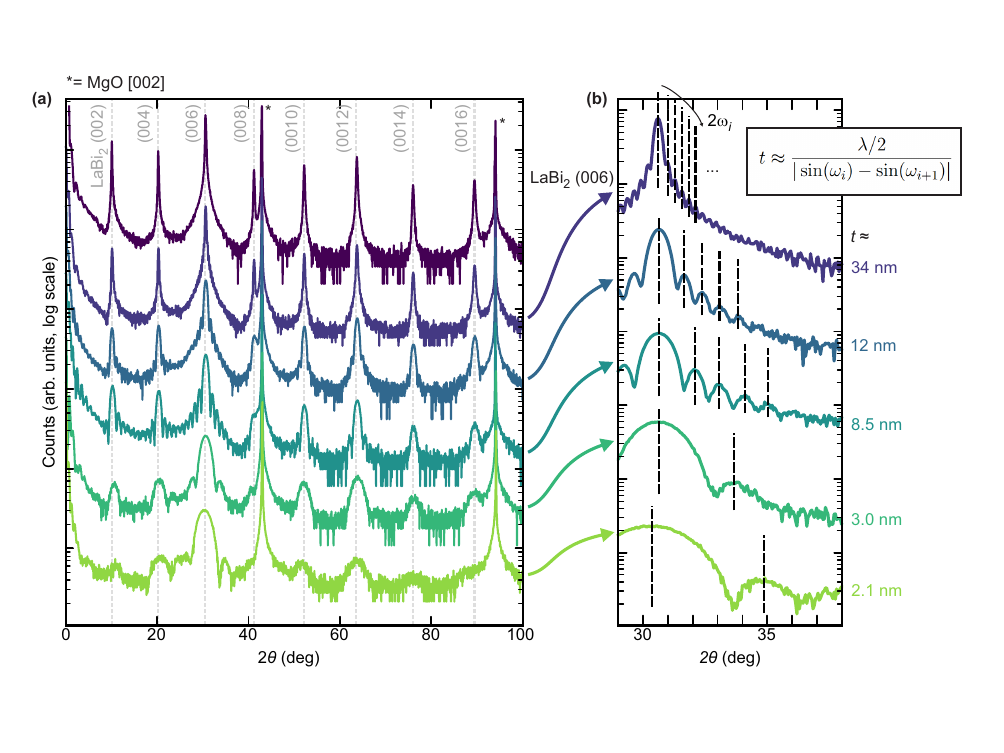}
    \caption{(a) Full X-ray diffraction patterns of single-phase \ce{LaBi2} films down to 2.1 nm in thickness. (b) Laue fringe analysis of (006) Bragg peak used to estimate each film's thickness. Here, $\lambda$ is X-ray wavelength. Note that this analysis is performed on both sides of the Bragg peak and the reported thickness is averaged over all observable Laue fringe pairs.}
    \label{FigS1}
\end{figure}
\newpage

\begin{figure}[h]
    \centering
    \includegraphics[width=170mm]{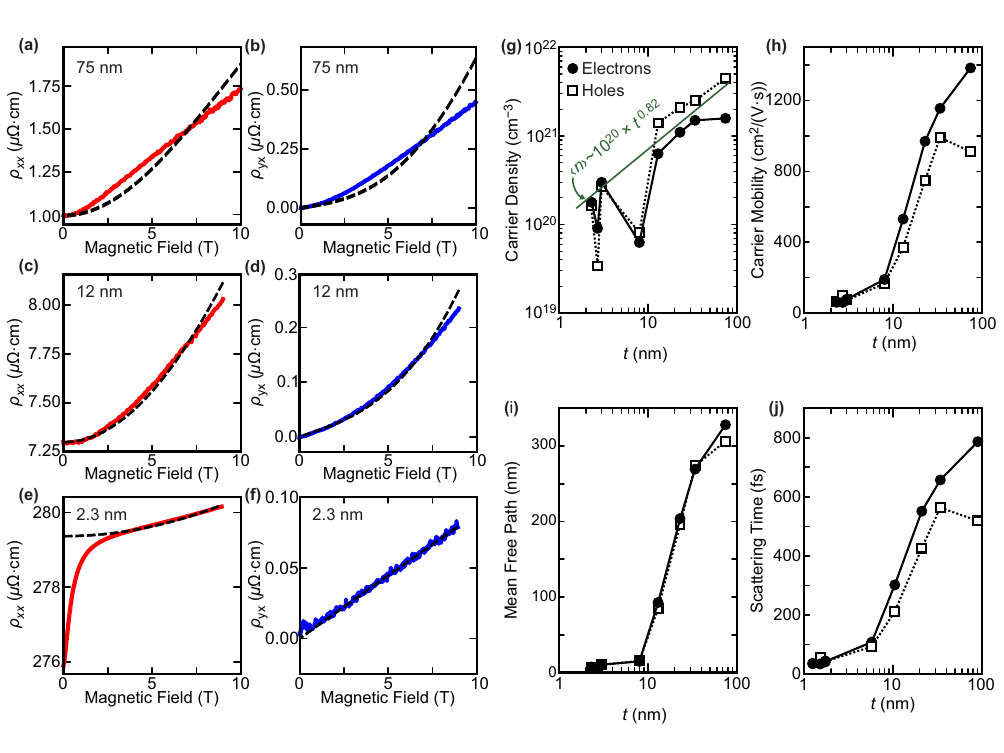}
    \caption{Normal-state sheet resistivity (a,c,e) and Hall resistivity (b,d,f) versus perpendicular magnetic field strength at $T = 2$~K for three representative thicknesses. Fits to the two-carrier Drude model are presented in black. The resulting (g) carrier densities and (h) mobilities are presented for both carrier species, as well as the (i) mean free paths and (j) transport scattering times estimated within the free electron model. A power-law fit of the species-averaged carrier density $<n>$ is shown in green in (g) which is used to estimate $p_\mathrm{F}$ for the sake of the KF framework.}
    \label{FigS2}
\end{figure}
\newpage

\begin{figure}[h]
    \centering
    \includegraphics[width=85mm]{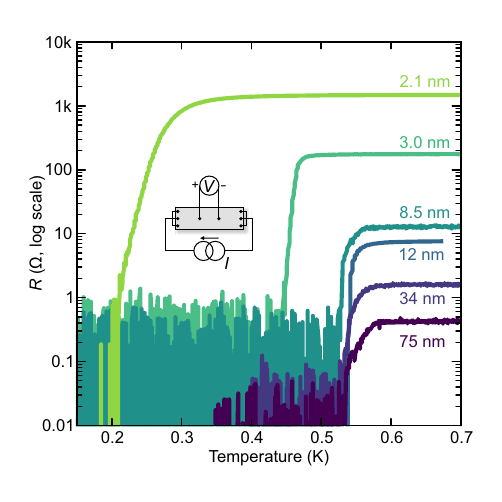}
    \caption{Raw-data versions of the superconducting transitions shown in Fig. 1 of the main text. The four-point resistance is plotted on a log scale for visibility of all samples and does not account for variations in sample geometry.}
    \label{FigS3}
\end{figure}
\newpage

\begin{figure}[h]
    \centering
    \includegraphics[width=170mm]{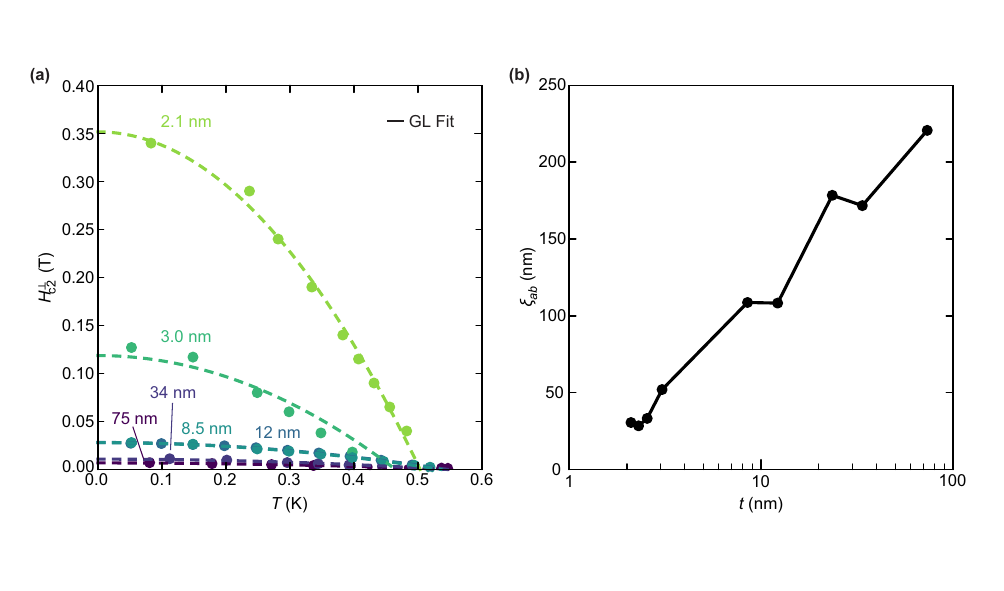}
    \caption{(a) Upper critical fields for $H$ perpendicular to the sample plane across the thickness series, fit to the single-gap Ginzberg-Landau (GL) formula $H_\mathrm{c2}^\perp$(T)$=(\Phi_0/2\pi\xi_{ab}^2)(1-(T_\mathrm{c}/T_{\mathrm{c},H=0})^2)$ where $\xi_{ab}$ is the superconducting coherence length. (b) $\xi_{ab}$ as estimated by the GL fitting versus film thickness.}
    \label{FigS4}
\end{figure}
\newpage

\begin{figure}[h]
    \centering
    \includegraphics[width=170mm]{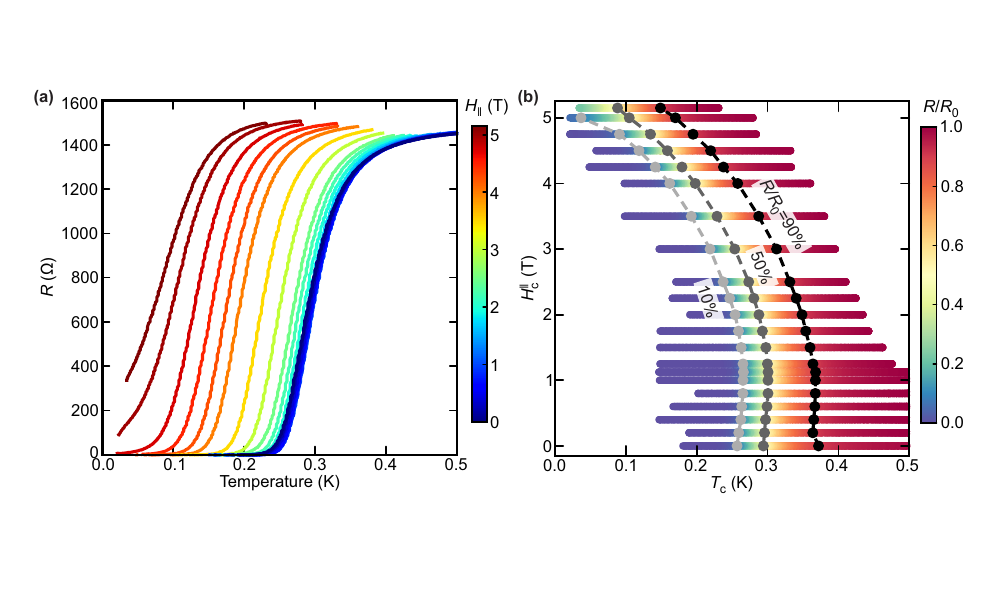}
    \caption{(a) Temperature-dependent superconducting transitions for the $2.1$ nm film at various parallel magnetic fields, showing the subtle field-enhanced $T_\mathrm{c}$ reported in the main text. (b) $H_\mathrm{c}^\parallel(T)$ curves for the field-enhanced sample according to $R=0.1\times R_0$, $R=0.5\times R_0$ (used in main text), and $R=0.9\times R_0$.}
    \label{FigS4}
\end{figure}
\newpage


\begin{figure}[t]
    \centering
    \includegraphics[width=170mm]{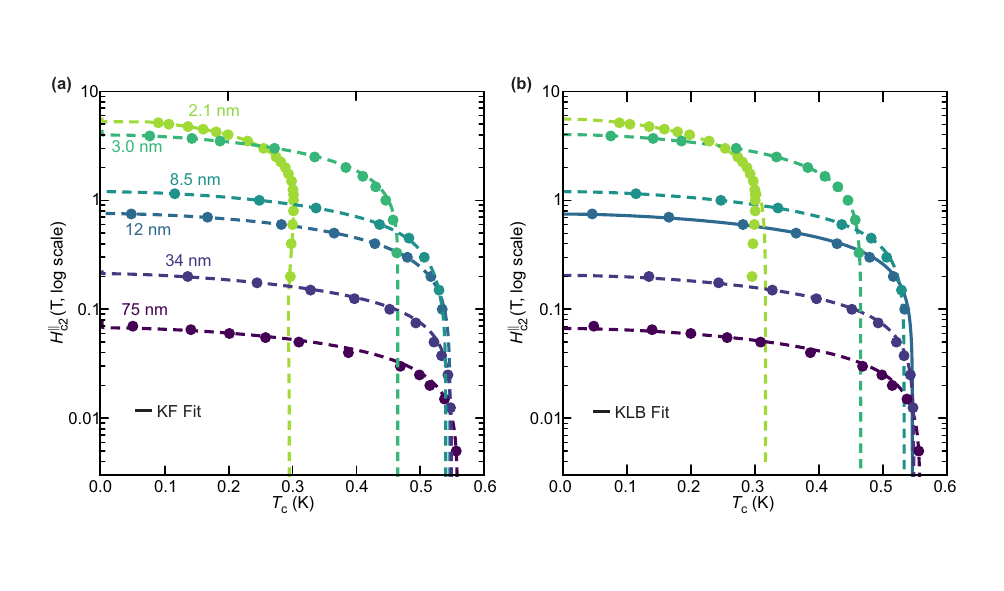}
    \caption{Full fitting curves of the (a) KF and (b) KLB models corresponding to the fit parameters reported in the main text. Here $H_\mathrm{c2}^\parallel(T)$ is plotted on a logarithmic scale in order to more clearly resolve the data across all thickness regimes.}
    \label{FigS5}
\end{figure}



\begin{table}[t]
\centering
\caption{Fit parameters for both KF and KLB models used throughout the text.}
\begin{tabular}{|c||c|c|c|c|c||c|c|}
\hline
\multicolumn{1}{|c||}{} 
& \multicolumn{5}{c||}{KF model} 
& \multicolumn{2}{c|}{KLB model} \\
\hline
$t$ (nm)
& $T_{\mathrm{c0}}$ (mK)
& $\tau_{\mathrm{S}}$ ($\times10^{-11}$~s)
& $\tau_{\mathrm{SO}}$ ($\times10^{-14}$~s)
& $\tau_{\mathrm{tr}}$ ($\times10^{-14}$~s)
& $\mathrm{sgn}(J)\zeta$
& $T_{\mathrm{c}}$ (mK)
& $\tau_{\mathrm{SO}}$ ($\times10^{-12}$~s) \\
\hline
2.1  & 352 & 5.49 & 0.99 & 0.27 & 26.8 & 317 & 0.10 \\
\hline
2.3  & 435 & 6.66 & 0.81 & 0.81 & 39.9 & 403 & 0.14 \\
\hline
2.7  & 483 & 6.86 & 0.97 & 0.96 & 36.8 & 453 & 0.17 \\
\hline
3.0  & 499 & 7.99 & 1.05 & 1.05 & 13.3 & 466 & 0.29 \\
\hline
8.5  & 582 & 6.68 & 2.54 & 1.42 & 39.9 & 534 & 3.68 \\
\hline
12   & 697 & 2.26 & 2.22 & 2.22 & 13.7 & 547 & 9.72 \\
\hline
23   & 647 & 2.88 & 1.92 & 1.79 & 31.2 & 535 & 60.5 \\
\hline
34   & 604 & 5.17 & 14.1 & 1.09 & 27.8 & 549 & 130 \\
\hline
75   & 600 & 6.64 & 1.67 & 1.67 & 13.3 & 558 & 1250 \\
\hline
\end{tabular}
\end{table}

\newpage

\begin{figure}[t]
    \centering
    \includegraphics[width=85mm]{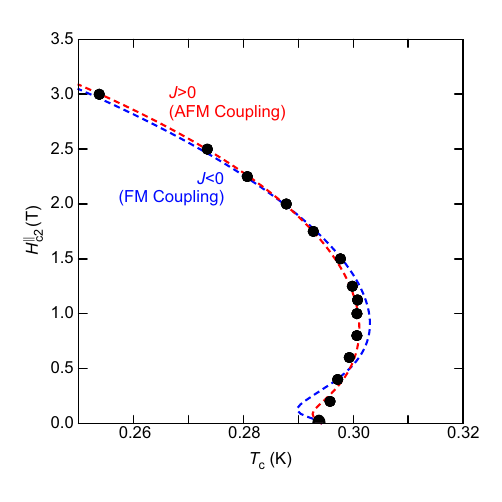}
    \caption{Near-$T_\mathrm{c}$ region of $H_\mathrm{c2}^\parallel(T)$ for the 2.1 nm film, comparing fit results of the KF model assuming antiferromagnetic ($J>0$, red) and ferromagnetic ($J<0$, blue) exchange coupling between impurity spins. The ferromagnetic exchange case is found to be less successful at reproducing the measured field-enhancement, hence our choice of $J>0$ throughout the work. See Table II for a summary of the fit parameters.}
    \label{FigS6}
\end{figure}

\begin{table}[t]
\centering
\begin{tabular}{|c|c||c|c|c|c|c|}
\hline
 $S$ 
 & $J$
 & $T_{\mathrm{c0}}$ (mK) 
 & $\tau_{\mathrm{S}}$ ($10^{-11}$ s)
 & $\tau_{\mathrm{SO}}$ ($10^{-14}$ s)
 & $\tau_{\mathrm{tr}}$ ($10^{-14}$ s)
 & $\mathrm{sgn}(J)\zeta$ \\
\hline
$5/2$&$>0$ 
& 352 & 5.49 & 0.81 & 0.27 & 26.8 \\
\hline
$5/2$& $<0$ 
& 497 & 1.60 & 0.045 & 0.013 & -53.7 \\
\hline
$3/2$& $>0$ 
& 333 & 7.87 & 0.962 & 0.027 & 40.2 \\
\hline
$1/2$& $>0$ 
& 317 & 13.2 & 0.895 & 0.025 & 72.9 \\
\hline
\end{tabular}
\caption{Impact of the choice of sgn$(J)$ and $S$ on KF fit parameters, focused on the 2.1 nm film. Note that for fixed sgn$(J)$ no obvious changes to the quality of the KF fit curves were observed for all values of $S$.}
\label{tab:spinexchange}
\end{table}


\newpage

\begin{figure}[h]
    \centering
    \includegraphics[width=170mm]{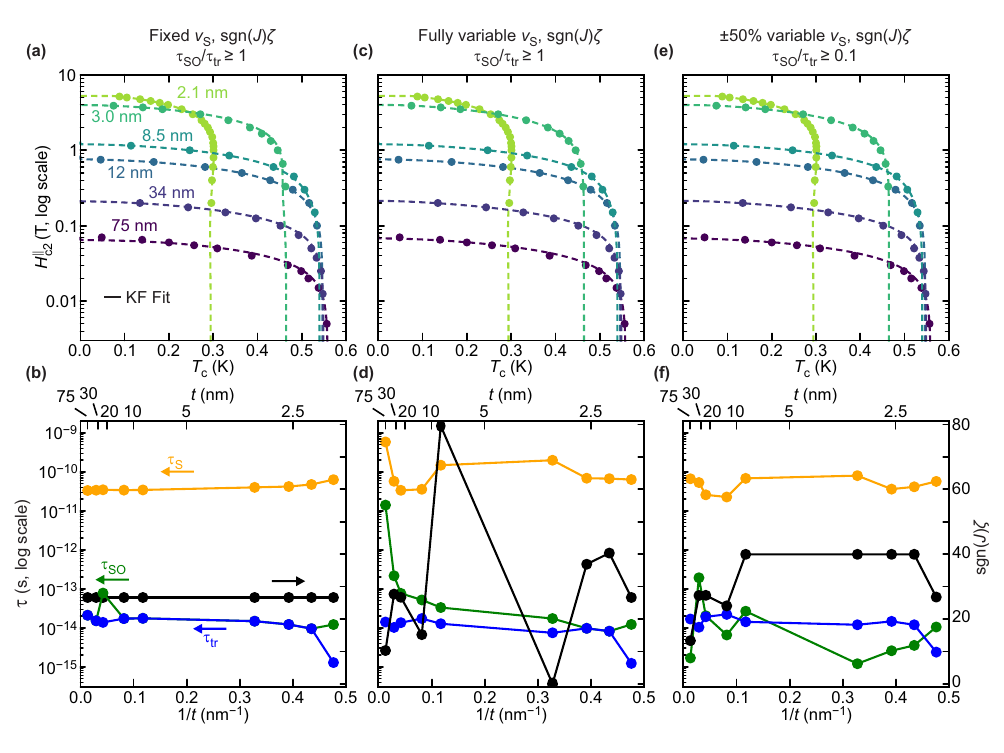}
    \caption{Comparison of KF fit approaches regarding the thickness-dependent treatment of $\nu_\mathrm{S}$ and sgn$(J)\zeta$ and constraints on $\tau_\mathrm{SO}$, each with little effect on the overall quality of the fits. (a) Holding both $\nu_\mathrm{S}$ and sgn($J$)$\zeta$ fixed to their ultrathin-limit value results in $\tau_\mathrm{SO}$ falling more often to the $\tau_\mathrm{SO}=\tau_\mathrm{tr}$ lower bound. (b) Allowing $\nu_\mathrm{S}$ and sgn($J$)$\zeta$ to vary fully leads to a sudden increase in $\tau_\mathrm{SO}$ and $\nu_\mathrm{S}$ when the orbital channel is most dominant in the thick-limit. (c) Reducing the lower bound on $\tau_\mathrm{SO}$ results in the occasional result $\tau_\mathrm{SO}/\tau_\mathrm{tr}<1$, which is unphysical.}
    \label{FigS7}
\end{figure}
\newpage


\end{document}